\documentclass[aps,prl,preprintnumbers,superscriptaddress,twocolumn]{revtex4-2}
\usepackage[dvips]{graphicx}
\usepackage{bm,latexsym,amsmath,amssymb,amsfonts,mathrsfs}
\usepackage{color}

\input{colordvi.tex}

\newcommand{\Amp}{\mathcal{M}}
\newcommand{\AmpF}{\mathcal{A}}
\newcommand{\Mpl}{M_{\rm Pl}}
\newcommand{\Ms}{M_{\rm s}}
\newcommand{\im}{{\rm Im}}

\newcommand{\nn}{\nonumber \\}

\begin{document}

\preprint{KOBE-COSMO-21-05, YITP-21-30}

\title{
Is the Standard Model in the Swampland? \\--- Consistency Requirements from Gravitational Scattering ---
}

\author{Katsuki Aoki}
\affiliation{Center for Gravitational Physics, Yukawa Institute for Theoretical Physics, Kyoto University, 606-8502, Kyoto, Japan}

\author{Tran Quang Loc}
\affiliation{Department of Applied Mathematics \& Theoretical Physics, University of Cambridge, Wilberforce Road, Cambridge CB3 0WA, United Kingdom}

\author{Toshifumi Noumi}
\affiliation{Department of Physics, Kobe University, Kobe 657-8501, Japan}

\author{Junsei Tokuda}
\affiliation{Department of Physics, Kobe University, Kobe 657-8501, Japan}

\begin{abstract}
We study compatibility of the Standard Model of particle physics and General Relativity by means of gravitational positivity bounds, which provide a necessary condition for a low-energy gravitational theory to be UV completable within the weakly coupled regime of gravity. In particular, we identify the cutoff scale of the Standard Model coupled to gravity
by studying consistency of light-by-light scattering. 
While the precise value depends on details of the Pomeron effects in QCD, the cutoff scale reads $10^{16}$GeV if the single-Pomeron exchange picture works well up to this scale. We also demonstrate that the cutoff scale is lowered to $10^{13}$GeV if we consider the electroweak theory without the QCD sector.
\end{abstract}

\maketitle

\paragraph{Introduction.---\!\!\!\!} When does General Relativity (GR) meet the Standard Model (SM) of particle physics? It is widely accepted to study gravitational interactions and particle interactions independently due to the large hierarchy between these forces; meanwhile, it is widely believed that all the interactions are ultimately unified by quantum gravity. If this is indeed the case, although the SM and GR are apparently independent at low energies, there must be hidden consistency relations between them by which we may extract information about quantum gravity from the well established physics, the SM and GR. 
\medskip

The Swampland Program~\cite{Vafa:2005ui} aims to clarify such consistency relations by studying necessary conditions for a low-energy gravitational effective field theory (EFT) to have a consistent ultraviolet (UV) completion.
A lesson there is that gravitational EFTs typically accommodate a cutoff scale well below the Planck scale. For example, if we consider a graviton-photon system, quantum gravity requires a charged state, otherwise the theory has a global $1$-form symmetry associated with a constant shift of the photon field~\cite{Banks:1988yz,Banks:2010zn,Harlow:2018tng}. The mass of the charged state specifies the cutoff scale of the original graviton-photon EFT, which is quantified by the Weak Gravity Conjecture as $m\leq \sqrt{2}|q|M_{\rm Pl}$~\cite{ArkaniHamed:2006dz}. It is well below the Planck scale $M_{\rm Pl}$ as long as the electric coupling $q$ is in the perturbative regime. See also review articles~\cite{Brennan:2017rbf,Palti:2019pca, vanBeest:2021lhn} for other related developments in the Swampland Program.

\medskip
It is well known that unitarity and analyticity of two-to-two scattering amplitudes lead to necessary conditions for a low-energy EFT to have a standard UV completion. In particular, the bounds on the Wilson coefficients are called the positivity bounds~\cite{Adams:2006sv}. While it has been a nontrivial issue how to derive rigorous bounds in the presence of gravity due to the $t$-channel graviton pole, recent works~\cite{Hamada:2018dde, Tokuda:2020mlf, Herrero-Valea:2020wxz} have clarified under which conditions (approximate) positivity bounds should hold
\footnote{See~\cite{Bellazzini:2019xts, Alberte:2020jsk, Arkani-Hamed:2020blm,Caron-Huot:2021rmr} for related discussions on positivity bounds in gravitational theories. Also, see~\cite{Cheung:2014ega,Andriolo:2018lvp,Hamada:2018dde,Chen:2019qvr,Bellazzini:2019xts,Loges:2019jzs,Andriolo:2020lul,Loges:2020trf,Aalsma:2020duv,Cremonini:2020smy,Alberte:2020bdz} for earlier applications of positivity bounds to the Swampland Program.}. 
The gravitational positivity bounds hold when gravity is UV completed in a weakly coupled way, realizing Regge behavior of high-energy scattering, which is indeed the case in perturbative string theory.

\medskip
Ref.~\cite{Alberte:2020bdz} studied the positivity bound on QED coupled to gravity, and predicted a cutoff scale $\Lambda\sim \sqrt{em_eM_{\rm Pl}} \sim 10^8$GeV in terms of the electron charge $e$ and the electron mass $m_e$, under several assumptions clarified shortly
\footnote{\label{footnote:AJNS}A similar cutoff $\Lambda\sim \sqrt{m_eM_{\rm Pl}/e}$ is implied from positivity in the presence of a hidden sector that is coupled to photon only through gravity~\cite{Andriolo:2018lvp}.}.
This value is much lower than the expected scales of unification and quantum gravity, while the analysis lacks other known physics, the electroweak (EW) and QCD sectors, and $10^8$GeV is not necessarily the scale of new physics. Also, $\Lambda\sim \sqrt{em_eM_{\rm Pl}}$ implies that a massless charged fermion is in the Swampland since we have $\Lambda\to0$ in the massless limit $m_e\to0$, which is somewhat surprising. 
In this letter, we identify an upper bound of the scale of new physics by completing the full SM analysis,
after reviewing gravitational positivity bounds and revisiting their implications for QED.

\medskip
\paragraph{Gravitational Positivity.---\!\!\!\!}
In this letter, we focus on the light-by-light scattering $\gamma\gamma\to\gamma\gamma$ in the SM coupled to GR, which has to be interpreted as a low-energy EFT of quantum gravity. The scattering amplitude is denoted by $\Amp(s,t)$, and $(s,t,u)$ are Mandelstam variables satisfying $s+t+u=0$. To manifest the $s\leftrightarrow u$ crossing symmetry, we consider the helicity sum,
\begin{eqnarray}
\Amp (s,t)&=&\Amp(1^+2^+3^+4^+)+\Amp(1^+2^-3^+4^-)
\nn
&&
+\, \Amp(1^-2^-3^-4^-)+\Amp(1^-2^+3^-4^+)
\,,
\end{eqnarray}
where $1,2$ are ingoing photons and $3,4$ are outgoing ones, and $\pm$ is the helicity.

\begin{figure}
\includegraphics[width=80mm]{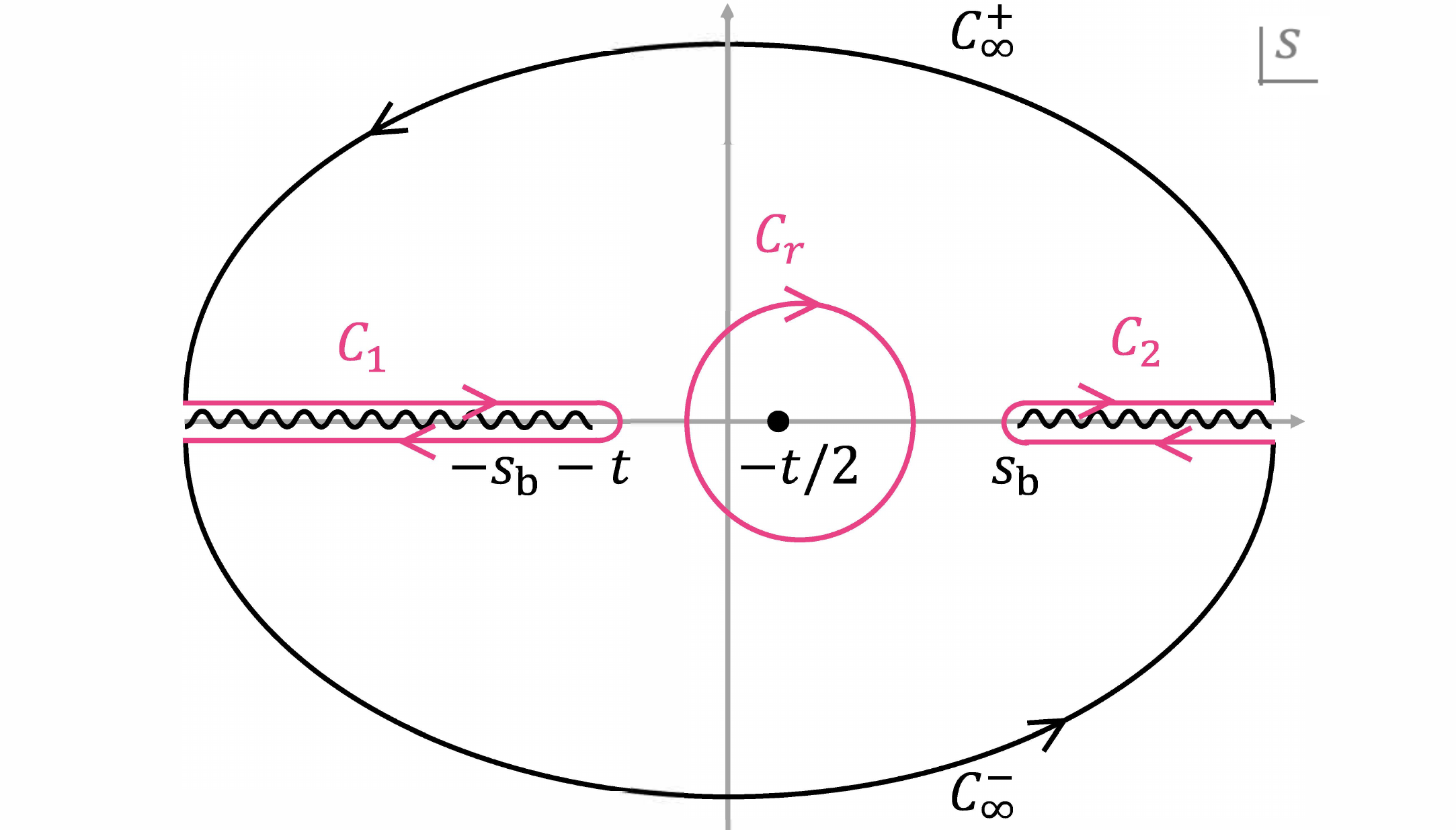}
\caption{Analytic structure of $\Amp(s,t)$ on the complex $s$-plane except for $s$- and $u$-channel poles, and the integration contour to derive the dispersion relation for fixed $t<0$ up to $\mathcal{O}(e^2/\Mpl^{2})$. The wavy line denotes branch cuts.}
\label{contour}
\end{figure}

\medskip
We assume several properties of $\Amp(s,t)$ on the complex $s$-plane for a fixed $t$ at least up to $\mathcal{O}(\Mpl^{-2})$: unitarity, analyticity, and a mild behavior in the Regge limit of the form $\lim_{|s|\to\infty}|\Amp(s,t<0)/s^2|\to0$
\footnote{These properties are indeed satisfied in known amplitudes in perturbative string theory. Furthermore, the analyticity and the mild behavior are inferred from causality and locality at least for a gapped system, so in the literature these properties are sometimes called `axioms'.}. 
We can then derive the twice-subtracted dispersion relation for a fixed $t<0$ by considering the integration contour shown in Fig.~\ref{contour}. 
The amplitude at the reference point $s=-(t/2)$ is expressed by integrals along the branch cuts ($\mathcal C_1 + \mathcal C_2$), the infinitely large semi-circles $(\mathcal C^+_\infty + \mathcal C^-_\infty)$, and the poles on the complex $s$-plane. Contributions from $\mathcal C^+_\infty + \mathcal C^-_\infty$ vanish thanks to the mild behavior in the Regge limit. 
We then consider the low-energy expansion, $\Amp=\sum_{n=0}^\infty\frac{c_n(t)}{n!}\left(s+\frac{t}{2}\right)^n+\frac{2(s^4+u^4)}{\Mpl^2 stu}$, where the second term manifests the graviton poles. In terms of 
$c_2(t)$, we have
\begin{equation}
\frac{c_2(t)}{2}-\frac{4}{\Mpl^2t}=\frac{2}{\pi}\int^\infty_{s_{\rm b}}\mathrm{d}s'\,\frac{\im\, \Amp(s'+i\epsilon,t)}{\left(s'+(t/2)\right)^3}\,,\label{eq:disp2}
\end{equation}
with using the $s\leftrightarrow u$ crossing symmetry. Here $s_{\rm b}=4m_e^2$ in our case, $m_e$ being the mass of the electron, the lightest charged particle. 
By definition, EFT can compute $\Amp$ up to the cutoff scale. Supposing EFT is valid at $s<\Lambda^2$, the quantity,
\begin{equation}
B^{(2)}(\Lambda,t):=c_2(t)-\frac{4}{\pi}\int^{\Lambda^2}_{s_{\rm b}}\mathrm{d}s'\,\frac{\im\,\Amp(s'+i\epsilon,t)}{(s'+(t/2))^3}\,,
\end{equation}
is calculable within EFT.
We emphasize that $B^{(2)}(\Lambda,t)$ is regular in the forward limit while $\Amp$ is not.
In terms of $B^{(2)}(\Lambda,t)$, eq.~\eqref{eq:disp2} reads~\cite{Bellazzini:2016xrt, deRham:2017imi}
\begin{equation}
B^{(2)}(\Lambda,t)-\frac{8}{\Mpl^2t}=\frac{4}{\pi}\int^\infty_{\Lambda^2}\mathrm{d}s'\,\frac{\im\,\Amp(s'+i\epsilon,t)}{(s'+(t/2))^3}\,.\label{eq:disp3}
\end{equation}
In the gravity decoupling limit $\Mpl\to\infty$, we can safely take the forward limit $t\to0$ in \eqref{eq:disp3} to get $B^{(2)}(\Lambda,0)>0$ thanks to the optical theorem~\cite{Adams:2006sv}. However, the forward limit of~\eqref{eq:disp3} is subtle in the presence of gravity because of the singular second term on the LHS. To obtain the finite expression of \eqref{eq:disp3} in the forward limit, one needs to see the cancellation of singular terms of both sides in eq.~\eqref{eq:disp3} and evaluate the $\mathcal{O}(t^{0})$ term carefully. Such computations have been explicitly done in~\cite{Tokuda:2020mlf} under the assumption of the Regge behavior
\begin{equation}
\im\,\Amp(s,t)=f(t)\left(\frac{s}{\Ms^2}\right)^{2+\alpha't+\alpha''t^2+\cdots} + \cdots
\,,\label{reg}
\end{equation}
at the UV regime, $s\gg\Ms^2(>\Lambda^2)$.
Here, $f(t)$ denotes a dimensionless function that is regular at $t=0$. The terms $\alpha'>0$ and $\alpha''$ are constants. Ellipses in the exponent stand for the higher-order terms in $t$ and we suppressed contributions from states which are irrelevant for Reggeizing the graviton exchange and the sub-leading terms of the Regge behavior.
The scales $\Ms$ and $\alpha'$ will be related to the mass scale of the physics which Reggeizes the amplitude. In string theory examples, the scattering amplitude exhibits the Regge behavior with $\alpha'\sim\Ms^{-2}$ via the string higher-spin states where $\Ms$ is the mass scale of the lightest higher-spin states, namely the string scale.
It is shown that~\cite{Tokuda:2020mlf}
\begin{equation}
B^{(2)}(\Lambda):=B^{(2)}(\Lambda,0)>-\mathcal{O}(\Mpl^{-2}\Ms^{-2})\,,\label{posi}
\end{equation}
assuming a single scaling $|(\partial_tf/f)_{t=0}|,|\alpha''/\alpha'|,\alpha'\lesssim \mathcal{O}(\Ms^{-2})$.
The precise value and the sign of the RHS will depend on the details of UV completion.
Although the small amount of negativity is still allowed, RHS is suppressed by not only $\Mpl^{-2}$ but also $\Ms^{-2}$ which is small enough to provide the constraints on the SM amplitudes with gravity
\footnote{The same order-estimate was performed in~\cite{Hamada:2018dde}, but the recent work~\cite{Tokuda:2020mlf} refined the bound more explicitly in terms of $f(t)$, $\alpha'$, and $\alpha''$. Also, more recently, the paper~\cite{Alberte:2020bdz} considers the gravitational positivity bounds without assuming the mild high-energy behavior in the Regge limit, and argues that the negativity of RHS may be allowed from the EFT perspective. In this case, however, the order of magnitude of RHS cannot be determined. }.

\begin{figure}[t]
\centering
 \includegraphics[width=1\linewidth]{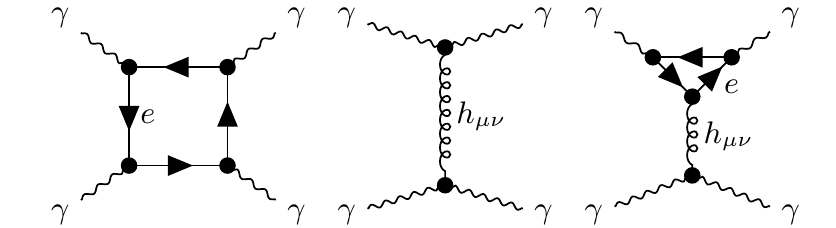}	
\caption{Feynman diagrams relevant for $\Amp_{\rm QED} $ and $\Amp_{\rm GR}$.}
\label{fig_QED}
\end{figure}

\begin{figure}[t]
\centering
\includegraphics[width=1\linewidth]{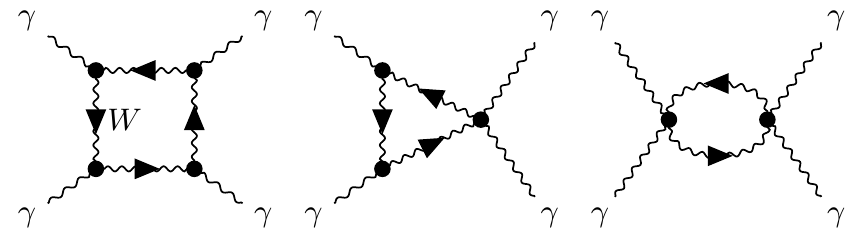}	
\caption{Feynman diagrams relevant for $\Amp_{\rm Weak} $.}
\label{fig_weak}
\end{figure}

\begin{figure}[t]
\centering
\includegraphics[width=1\linewidth]{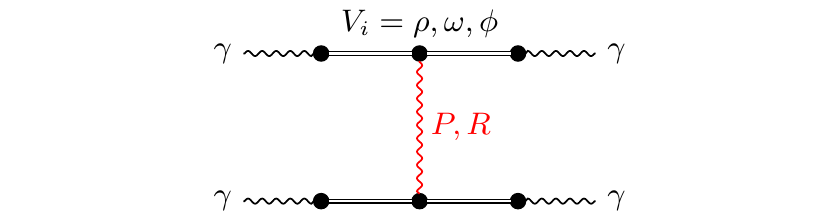}	
\caption{Feynman diagrams relevant for $\Amp_{\rm QCD} $.}
\label{fig_QCD}
\end{figure}

\medskip
In summary, the general properties of the amplitudes lead to the bound \eqref{posi} as a consistency condition, where $B^{(2)}(\Lambda)$ is computed by the EFT, the SM coupled to GR in the present case. The amplitude for the light-by-light scattering at $s<\Lambda^2$ can be decomposed as 
\begin{equation}
\label{amp_def}
\Amp(s,t)=\Amp_{\rm QED}+\Amp_{\rm Weak}+\Amp_{\rm QCD} +\Amp_{\rm GR}
\,,
\end{equation}
where relevant diagrams for each sector are given by Figs.~\ref{fig_QED}-\ref{fig_QCD} and their crossed versions. The corresponding $B^{(2)}_i(\Lambda)$ with $i=$ QED, Weak, QCD, and GR are computed accordingly. 
In general, EFT must contain higher derivative operators representing corrections from UV physics, which will be taken care of in the following discussion.
As we will see, they turn out to be irrelevant for our purpose except for the QED case.

\medskip
Because the SM is renormalizable, the SM amplitude satisfies the twice-subtracted dispersion relation, giving
\begin{equation}
B^{(2)}_i(\Lambda)=\frac{4}{\pi}\int_{\Lambda^2}^{\infty} \mathrm{d}s'\,\frac{\im\,\AmpF_i(s'+i\epsilon)}{s'^3}\,,
\label{B_inf}
\end{equation}
for $i=$ QED, Weak, and QCD. Here, $\AmpF_i(s):=\Amp_i(s,t=0)$ is the forward limit amplitude. The relations \eqref{B_inf} conclude $B^{(2)}_{i} (\Lambda \to \infty) =0$. The same relation does not need to apply to the GR sector, however, because GR is not UV complete; as we will see, $B^{(2)}_{\rm GR}(\Lambda \to \infty) \to  {\rm constant}<0$. As $\Lambda$ increases, i.e., as~the SM coupled to GR is extrapolated to high-energy scales, the GR contribution eventually dominates over the SM contributions, leading to violation of \eqref{posi}. The maximum cutoff scale of the SM coupled to GR is determined when the inequality \eqref{posi} is saturated.

\medskip
\paragraph{Positivity in QED.---\!\!\!\!}
We begin with the light-by-light scattering in QED coupled to GR which was discussed recently in~\cite{Alberte:2020bdz}.
The leading QED contribution is the one-loop diagram shown in the first diagram of Fig.~\ref{fig_QED}. The forward limit amplitude is given by
\begin{equation}
\AmpF_{\rm QED}  \approx -8\alpha^2\left(6+\ln^2\frac{m_e^2}{-s}+2\ln\frac{m_e^2}{-s}  \right)  + (s\leftrightarrow -s)
\label{A_QED}
\end{equation}
in the high-energy limit $|s| \gg m_e^2$, where $\alpha=e^2/4\pi$ is the fine-structure constant.
$B^{(2)}(\Lambda)$ from the QED process at one-loop level is
\begin{equation}
B^{(2)}_{\rm QED}\approx \frac{64\alpha^2}{\Lambda^4}\left(\ln \frac{\Lambda}{m_e} - \frac{1}{4} \right)
\label{B_QED}
\end{equation}
for $\Lambda \gg m_e$. 
Regarding $\Amp_{\rm GR}$, we have the tree and one-loop diagrams up to $\mathcal{O}(\Mpl^{-2})$ as shown in Fig.~\ref{fig_QED}. The tree level (pole) contribution is cancelled with the high-energy integral of RHS of~\eqref{eq:disp3}. As a result, the one-loop diagram is the leading gravitational contribution to the bound \eqref{posi}. In the high-energy limit, we obtain
\begin{equation}
B^{(2)}_{\rm GR} \approx -\frac{22 \alpha}{45\pi m_e^2 \Mpl^2}
\,.
\label{B_GR}
\end{equation}
Whereas $B^{(2)}_{\rm QED}$ vanishes as $\Lambda \to \infty$, $B^{(2)}_{\rm GR}$ approaches a negative constant in the limit $\Lambda \to \infty$. This is why we obtain a nontrivial cutoff scale from (approximate) positivity bounds.

\medskip
It is also convenient to remark that the result~\eqref{B_GR} can be used even in the later analysis beyond QED. In general, the one-loop contribution to $B^{(2)}_{\rm GR}$ from charged particles should be proportional to $e^2/\Mpl^2$ and the dimensional analysis concludes $B^{(2)}_{\rm GR} \propto e^2/(m^2 \Mpl^2)$, where $m$ is the mass of the propagating particle in the loop. Therefore, the lightest charged particle should provide the dominant contribution to $B^{(2)}_{\rm GR}$. We thus take into account the electron loop only to compute $B^{(2)}_{\rm GR}$ throughout this letter. 

\medskip
Now we discuss implications of the gravitational positivity bound~\eqref{posi}. 
First, we can simply discard the $\mathcal{O}(\Mpl^{-2}\Ms^{-2})$ uncertainty in \eqref{posi} because the GR contribution \eqref{B_GR} is much larger than the uncertainty.
Second, as we mentioned earlier, there are potential higher derivative corrections that are originated from UV physics above the scale $\Lambda$. From the EFT perspective, its contribution to $B^{(2)}$ can be estimated as
$
B^{(2)}_{\rm UV}=\frac{\alpha_{\rm UV}}{\Lambda^4},
$
where the dimensionless parameter $\alpha_{\rm UV}$ characterizes the size of interactions at the scale $\Lambda$ and satisfies $|\alpha_{\rm UV}|\lesssim 1$.

\medskip
All in all, the gravitational positivity implies the bound $B^{(2)}_{\rm QED}+B^{(2)}_{\rm UV}+B^{(2)}_{\rm GR}>0$, yielding
\begin{equation}
\frac{64\alpha^2}{\Lambda^4}\left(\ln \frac{\Lambda}{m_e} - \frac{1}{4} \right)
+\frac{\alpha_{\rm UV}}{\Lambda^4}>\frac{22 \alpha}{45\pi m_e^2 \Mpl^2}\,.
\end{equation}
If the first term is dominant, we have the same bound as~\cite{Alberte:2020bdz}, $\Lambda\lesssim \sqrt{em_eM_{\rm Pl}}$. On the other hand, if the second term is dominant and $\alpha_{\rm UV}\sim1$, we find $\Lambda\lesssim \sqrt{m_eM_{\rm Pl}/e}$, which is the one obtained in \cite {Andriolo:2018lvp} from a slightly different setup (see~\cite{Note2}). In either case, the typical size is
$\Lambda<\Lambda_{\rm QED}\sim 10^{8}{\rm GeV}$
which is regarded as the maximum cutoff scale of QED coupled to GR. A new physics is required below $\Lambda_{\rm QED}$ to satisfy the bound \eqref{posi}. Needless to say, we already know the ``new" physics, weak force and strong force, in nature and these physics contribute to the light-by-light scattering well below $10^8$GeV.

\begin{figure}[t] 
\centering
 \includegraphics[width=0.9\linewidth]{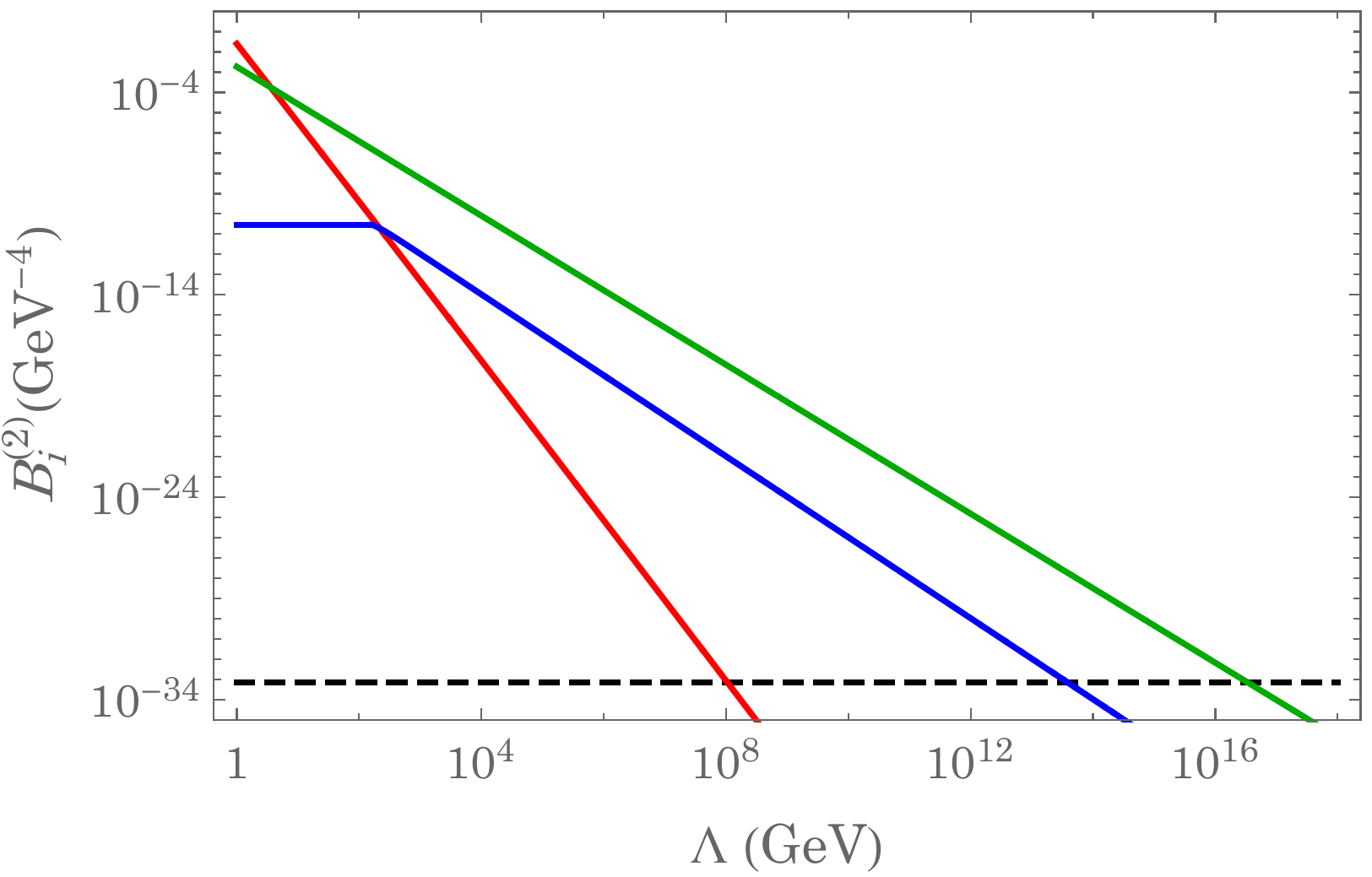}	
 \caption{The $\Lambda$ dependence of $B^{(2)}_i$ where $i=$ QED (red), Weak (blue), and QCD (green), and the black dashed line represents $-B^{(2)}_{\rm GR}$. The intersection between the solid line and the dashed line determines the cutoff $\Lambda_i$. }
 \label{fig_all_B}
\end{figure}

\medskip
\paragraph{Positivity in Electroweak Theory.---\!\!\!\!}
We then include the weak sector into our consideration. While charged lepton loops provide the same contribution as \eqref{B_QED} (after a replacement of $m_e$ by the lepton masses), W bosons yield a qualitatively different contribution because of the spin-1 nature. 
In the high-energy limit $(|s|\gg m_W^2)$, the one-loop amplitude is
\footnote{The one-loop diagrams are calculated by using the Mathematica packages {\sc FeynArts}~\cite{Hahn:2000kx} and {\sc FeynCalc}~\cite{Shtabovenko:2020gxv}, and the loop integrals are evaluated by {\sc Package-X}~\cite{Patel:2015tea}. As a consistency check, we confirm the desired crossing symmetries, the relation \eqref{B_inf}, and the agreement with two different gauge choices, the Feynman-'t Hooft gauge and the unitary gauge.}
\begin{equation}
\AmpF_{\rm Weak} \approx \frac{32\alpha^2 }{m_W^2} s \ln \frac{m_W^2}{-s} + (s\leftrightarrow -s)
.
\label{A_Weak}
\end{equation}
In contrast to \eqref{A_QED}, the imaginary part of the amplitude grows linearly in $s$ in the high-energy limit. Accordingly, the weak sector contribution to $B^{(2)}$ reads
\begin{equation}
B^{(2)}_{\rm Weak} \approx  \frac{128\alpha^2}{m_W^2 \Lambda^2}
\,,
\label{B_Weak}
\end{equation}
which decreases as $\Lambda^{-2}$. Then, the W boson contribution $B^{(2)}_{\rm Weak}$ eventually dominates over the fermion loop contributions~\eqref{B_QED} at UV (see Fig.~\ref{fig_all_B}, where we plot $B^{(2)}_i$ without using the high-energy approximation). The UV physics effect $B^{(2)}_{\rm UV}\propto \Lambda^{-4}$ also becomes subdominant in the same regime.
As a result, we obtain the cutoff which is much larger than the one obtained in QED case,
\begin{equation}
\Lambda_{\rm EW}=\sqrt{\frac{2880 \pi \alpha}{11}} \frac{m_e \Mpl}{m_W} \simeq 3.8 \times 10^{13} {\rm GeV}\,.
\end{equation}

\medskip
It is worth mentioning that after taking the high-energy limit $\Lambda \gg m$, the fermion contribution \eqref{B_QED} is almost independent of the fermion mass and the mass of spin-1 particle (W boson) appears in the denominator of \eqref{B_Weak}. Therefore, we may continue to increase $\Lambda$ even if new charged spin-1/2 or spin-1 states, namely new physics, appear because they are subdominant in $B^{(2)}(\Lambda)$. The result must be insensitive to inclusion of new charged particles at UV regime as far as the theory is weakly coupled
\footnote{The inclusion of a charged spin-0 particle does not change the situation as well. See~\cite{Alberte:2020bdz} for the analysis in scalar QED.}. 
On the other hand, QCD is not a weakly coupled theory and, more importantly, QCD accommodates mesons that are lighter than W bosons. The result here must be insensitive to unknown UV physics involving up to spin-1 particles but sensitive to QCD.

\medskip
\paragraph{Positivity in Standard Model.---\!\!\!\!}
We finally take into account all the known physics and evaluate the cutoff scale of the SM by means of the gravitational positivity bounds. 
Since (non-gravitational) QCD amplitudes have to satisfy \eqref{B_inf}, we can compute $B^{(2)}_{\rm QCD}$ from the imaginary part of the forward limit amplitude $\im\,\AmpF_{\rm QCD}$ at UV. A nontriviality here is that in the forward limit, the momentum transfer is soft and so the non-perturbative physics of QCD contributes to $\im\,\AmpF_{\rm QCD}$ even at UV via $t$-channel diagrams.
To compute the light-by-light scattering in the forward limit, we use the vector meson dominance model (VDM) and consider intermediate hadronic excitations, which we call the VDM-Regge model following~\cite{Klusek-Gawenda:2016euz}.

\medskip
The relevant Feynman diagrams in the VDM-Regge model are shown in Fig.~\ref{fig_QCD}. The photon is supposed to transform into vector mesons $V_i=\rho,\omega,\phi$ before the collision and the mesons undergo the hadronic processes exchanging Pomeron and Reggeon ($P$ and $R$ in Fig.~\ref{fig_QCD}). 
The corresponding amplitude reads~\cite{Klusek-Gawenda:2016euz}
\begin{equation}
\Amp_{\rm QCD}\approx 4\left( \sum_i C^2_{\gamma \to V_i} \right)^2 \Amp_{VV\to VV} \left( \sum_j C^2_{V_j \to \gamma } \right)^2,
\end{equation}
where $C^2_{\gamma \to V_i}$ are the transition constants and the hadronic interactions are supposed to be the universal form.
$\Amp_{VV\to VV}$ is composed of two contributions, the Pomeron exchange and the Reggeon exchange, where the former one provides the faster than linear growth in $s$ while the latter one is subdominant at UV.
Also, the prefactor $4$ originates from the helicity sum.
The imaginary part of the amplitude reads
\footnote{We use the same references cited by~\cite{Klusek-Gawenda:2016euz}. First, the transition constants are $C^2_{\gamma \to \rho}=\frac{\alpha}{2.54}, ~ C^2_{\gamma \to \omega}=\frac{\alpha}{20.5}, ~ C^2_{\gamma \to \phi}=\frac{\alpha}{11.7}$, taken from~\cite{Ioffe:1985ep}. Adopting \cite{Donnachie:1992ny}, the Pomeron exchange contribution grows as $s^{1.08}$ and the overall factor is chosen such that ${\rm Im}\AmpF_{VV\to VV}=8.56 \,{\rm mb}\,s(s/s_0)^{0.08}$~\cite{Klusek:2009yi}. }
\begin{equation}
\label{QCD_amp}
{\rm Im} \AmpF_{\rm QCD} \approx 
25\alpha^2 \frac{s}{{\rm GeV}^2} \left(\frac{s}{s_0}\right)^{0.08}
\end{equation}
for $s\gg {\rm GeV}^2$, where we introduced $s_0 \sim {\rm GeV}^2$.

\medskip
Then, it is straightforward to calculate $B^{(2)}_{\rm QCD}$ using \eqref{B_inf} and the $s\leftrightarrow u$ symmetry.
All $B^{(2)}_i(\Lambda)~(i=$ QED, Weak, and QCD) are shown in Fig.~\ref{fig_all_B}, where the dashed line is $-B^{(2)}_{\rm GR}$. Since the QCD contribution dominates over $B^{(2)}_{\rm EW}=B^{(2)}_{\rm QED}+B^{(2)}_{\rm Weak}$ and also $B^{(2)}_{\rm UV}\propto \Lambda^{-4}$, the maximum cutoff scale of SM is determined by $B^{(2)}_{\rm QCD}+B^{(2)}_{\rm GR}=0$, yielding
\begin{equation}
\Lambda_{\rm SM}\simeq 3\times 10^{16} {\rm GeV}.
\end{equation}
This is one of our main results.

\medskip
A remark is needed before making a conclusion. Although the VDM-Regge model here assumes that the single Pomeron exchange captures the scattering process well, 
it is not clear up to which scale the model is trustable since there is no experimental input at such high-energy scales. 
To argue model-(in)dependence of our conclusion, we illustrate the following two cases that have the same value as \eqref{QCD_amp} at the GeV scale:
\begin{eqnarray}
{\rm Im}\AmpF_{\rm QCD} \approx 
\begin{cases} 
25\alpha^2 \frac{s}{{\rm GeV}^2} &\!\!{\rm (linear~growth)} , \\
25\alpha^2 \frac{s}{{\rm GeV}^2}\ln^2s/s_0   &\!\!{\rm (Froissart~type)} ,
\end{cases}
~~
\end{eqnarray}
where the linear growth in the former ansatz corresponds to a constant cross section while the second ansatz is motivated by the Froissart bound~\cite{Froissart:1961ux}.
The corresponding cutoff scale reads
\begin{eqnarray}
\Lambda_{\rm SM} \simeq 
\begin{cases} 
2\times 10^{15} {\rm GeV} &{\rm (linear~growth)}, \\
1\times 10^{17} {\rm GeV} &{\rm (Froissart~type)}.
\end{cases}
\end{eqnarray}
We conclude that the potential theoretical uncertainty in our calculations does not change $\Lambda_{\rm SM}$ drastically.

\medskip
\paragraph{Conclusion.---\!\!\!\!}
In this letter, we identified the cutoff scale of the Standard Model coupled to gravity as $10^{16}$GeV, applying gravitational positivity bounds to the light-by-light scattering $(\gamma\gamma\to\gamma\gamma)$. This means that quantum gravity requires a new physics below $10^{16}$GeV, otherwise the Standard Model falls into the Swampland. As we mentioned, weakly coupled charged particles up to spin-1 do not help to push up the cutoff scale, suggesting that beyond SM physics (described within non-gravitational QFT) at $E \gg {\rm GeV}$ would be irrelevant to our analysis.
A natural expectation would be thus that quantum gravity shows up around or below the obtained cutoff scale to reconcile the gravitational positivity. It is suggestive that this scale is close to the Grand Unification scale and the typical string scale. Nevertheless, it is worth again emphasizing that our result $\Lambda_{\rm SM} \sim 10^{16}$GeV is obtained from the consistency of the scattering amplitude based on the well established physics, the Standard Model and General Relativity
\footnote{Also note that we have assumed a weakly coupled UV completion where the graviton is Reggeized below the Planck scale. Our precise statement is if the SM coupled to GR is UV completed at a scale below the Planck scale, the scale should be less than $10^{16}{\rm GeV}$. It would be interesting to generalize our argument to strongly coupled UV completion of gravity. For this, one would need to first carefully reconsider the standard assumptions of positivity bounds such as locality and unitarity because super-Planckian physics such as black hole creation cannot be ignored.}.
Also, it is interesting that the Pomeron physics is crucial to understanding the 
cutoff scale of the Standard Model in light of the light-by-light scattering which is also an interesting process experimentally~\cite{Aaboud:2017bwk}.

\medskip
We also studied the electroweak theory without the QCD sector which may provide insights into the Swampland Program. 
The electroweak bound reads $\frac{m_W}{M_{\rm Pl}}=\sqrt{\frac{2880 \pi \alpha}{11}} \frac{m_e}{\Lambda_{\rm EW}}$, meaning that the $W$ boson mass is correlated with the electron mass for a given $\Lambda_{\rm EW}$. In particular, the massless limit of the electron requires the simultaneous massless limit of $W$ boson for a finite $\Lambda_{\rm EW}$.
This is reminiscent of D-brane realization of the Higgs mechanism in string theory~\cite{Cremades:2002cs}, where both of the electron mass and the $W$ boson mass are controlled by separation of D-branes.
Also, the electroweak bound yields a condition on the electron Yukawa coupling $y_e$ and the Weinberg angle $\theta_\text{W}$,
$
y_e\sin\theta_\text{W}=\sqrt{\frac{11}{1440}}
\frac{\Lambda_\text{EW}}{\Mpl}\,,
$
suggesting the existence of new swampland conditions on the coupling strengths. It would be interesting to explore these directions further from both top-down and bottom-up considerations to find quantitative predictions of quantum gravity.


\medskip
\begin{acknowledgments}
\paragraph{Acknowledgments.}
We would like to thank Yuji Yamazaki for useful discussions. The work of K.A. was supported in part by Grants-in-Aid from the Scientific Research Fund of the Japan Society for the Promotion of Science, No.~19J00895 and No.~20K14468.
T.N. is supported in part by JSPS KAKENHI Grant Numbers JP17H02894 and 20H01902, and MEXT KAKENHI Grant No.~21H00075.
J.T. was supported in part by JSPS Grants-in-Aid for Scientific Rersearch No. 202000912 and No.~21K13922.
\end{acknowledgments}

\bibliography{SMref}{}

\end{document}